\documentclass[10pt,conference]{IEEEtran}
\usepackage{epsfig,graphicx,subfigure,psfrag,amsmath,cases,bm}
\usepackage{latexsym,amssymb,amsmath,epsfig,subfigure,algorithm,mathtools}
\usepackage{algorithmic}
\usepackage{color}
\usepackage{url}
\usepackage{scrtime}
\usepackage{stfloats}
\usepackage[left=0.6in,top=0.70in,bottom=0.9in,right=0.6in]{geometry}
\author{\IEEEauthorblockN {Dongfang Xu\IEEEauthorrefmark {2}, Xianghao Yu\IEEEauthorrefmark {2}, Yan Sun\IEEEauthorrefmark {2}, Derrick Wing Kwan Ng\IEEEauthorrefmark {4}, and Robert Schober\IEEEauthorrefmark {2}}

\IEEEauthorrefmark {2}Friedrich-Alexander-Universit\"at
Erlangen-N\"urnberg, Germany, \IEEEauthorrefmark {4}The University
of New South Wales, Australia\vspace*{-4mm}

}

\newtheorem{T-Prob}{Transformed Problem}

\newcounter{TempEqCnt}

\DeclareMathOperator{\maxo}{maximize}
\DeclareMathOperator{\mino}{minimize}

 \newcommand{\qed}{\hfill \ensuremath{\blacksquare}}

\title{\vspace*{-1mm}Resource Allocation for Secure IRS-assisted Multiuser MISO Systems\vspace*{-5mm}}

\begin{document}
\maketitle
\vspace*{-3mm}
\begin{abstract}
In this paper, we study resource allocation design for secure communication in intelligent reflecting surface (IRS)-assisted multiuser multiple-input single-output (MISO) communication systems. To enhance physical layer security, artificial noise (AN) is transmitted from the base station (BS) to deliberately impair the channel of an eavesdropper. In particular, we jointly optimize the phase shift matrix at the IRS and the beamforming vectors and AN covariance matrix at the BS for maximization of the system sum secrecy rate. To handle the resulting non-convex optimization problem, we develop an efficient suboptimal algorithm based on alternating optimization, successive convex approximation, semidefinite relaxation, and manifold optimization. Our simulation results reveal that the proposed scheme substantially improves the system sum secrecy rate compared to two baseline schemes.
\end{abstract}
\vspace*{-1mm}
\section{Introduction}
\vspace*{-1mm}
Recently, intelligent reflecting surface (IRS)-assisted wireless communication systems have received considerable attention as a promising approach for providing cost-effective and power-efficient high data-rate communication services for the fifth-generation and beyond wireless communication systems \cite{di2019smart}\nocite{hu2018beyond,wu2018intelligent,8683145,yu2019miso,najafi2019intelligent}--\cite{qingqing2019towards}. Consisting of a set of small reflecting elements, IRSs can be easily and flexibly deployed on building facades and interior walls, improving communication service coverage \cite{di2019smart}. Compared to conventional relays and distributed antenna systems \cite{5490978}, passive reflectors embedded in IRSs require little operational power which makes them suitable for deployment in energy-constrained systems. Furthermore, due to their programmability and reconfigurability, IRSs can be adjusted on-demand such that a favourable radio propagation environment is obtained to improve system performance \cite{di2019smart}. As a result, several initial works have addressed technical issues regarding the design of IRS-assisted communication systems. For instance, the authors in \cite{wu2018intelligent} investigated the joint transmit beamforming and phase shift matrix design for maximization of the total received power of the user of an IRS-enhanced single-user system. In \cite{yu2019miso}, two computationally efficient suboptimal algorithms were developed for maximization of the spectral efficiency achieved by an IRS-assisted multiple-input single-output (MISO) communication system. However, these works did not consider security and the obtained results may not be applicable to systems where communication security is a concern.
\par
Recently, physical layer security has emerged as a promising technology to facilitate secure communication in wireless systems \cite{8349956}. By configuring multiple antennas at the base station (BS), beamforming can be employed to degrade the channel quality of eavesdroppers. In \cite{7222458}, a transmit beamforming algorithm was designed to achieve communication secrecy in a MISO wireless system. Furthermore, the authors of \cite{yu2019enabling} proposed two algorithms to maximize the secrecy rate in an IRS-assisted MISO wireless system. In \cite{shen2019secrecy}, the authors jointly optimized the beamforming vectors at the BS and the phase shifts at the IRS for maximization of the secrecy rate of a legitimate user. However, in \cite{yu2019enabling} and \cite{shen2019secrecy}, artificial noise (AN) is not employed for security enhancement. Nevertheless, AN transmission is an effective approach to improve physical layer security \cite{7955970}. Moreover, \cite{yu2019enabling} and \cite{shen2019secrecy} focused on the case of maximizing the secrecy rate of a single user and the proposed schemes may not be able to guarantee secure communication for multiuser IRS-assisted systems. The authors of \cite{chen2019intelligent} investigated the resource allocation algorithm design for maximization of the minimum secrecy rate among several legitimate users of an IRS-assisted multiuser MISO system. However, in \cite{chen2019intelligent} the unit modulus constraint introduced by the reflectors of the IRS was approximated by a convex constraint, which simplifies the optimization problem considerably and may lead to a performance loss. Therefore, the design of efficient resource allocation algorithms for maximization of the sum secrecy rate of IRS-assisted multiuser communication systems employing AN to impair eavesdroppers and imposing a unit modulus constraint for the IRS reflectors remains an open issue.
\par
Motivated by the above discussions, in this paper,  we  investigate the joint design of the phase shift matrix at the IRS and the downlink (DL) beamforming vectors and the AN covariance matrix at the BS for maximizing the system sum secrecy rate.
\vspace*{-1mm}
\section{System Model}
\vspace*{-1mm}
In this section, after introducing the notations used in this paper, we present the system model adopted for IRS-assisted communication.
\vspace*{-2mm}
\subsection{Notations}
\vspace*{-1mm}
In this paper, we use boldface capital and lower case letters to represent matrices and vectors, respectively. $\mathbb{R}^{N\times M}$ and $\mathbb{C}^{N\times M}$ denote the space of $N\times M$ real-valued and complex-valued matrices, respectively. $\mathbb{H}^{N}$ denotes the set of all $N$-dimensional complex Hermitian matrices. $\mathbf{I}_{N}$ indicates an $N\times N$ identity matrix. $|\cdot|$ and $||\cdot||_2$ denote the absolute value of a complex scalar and the $l_2$-norm of a vector, respectively. $\mathbf{x}^T$, and $\mathbf{x}^H$ stand for the transpose and the conjugate transpose of vector $\mathbf{x}$, respectively. $\mathbf{A}\succeq\mathbf{0}$ indicates that $\mathbf{A}$ is a positive semidefinite matrix. $\mathrm{Rank}(\mathbf{A})$, $\mathrm{Tr}(\mathbf{A})$, and $\left [ \mathbf{A} \right ]_{i,i}$ denote the rank, the trace, and the $(i,i)$-entry of matrix $\mathbf{A}$, respectively. $x_i$ denotes the $i$-th element of vector $\mathbf{x}$. $\mathrm{diag}(\mathbf{x})$ represents the $N\times N$ diagonal matrix with diagonal elements $x_1, \cdots, x_N$. unt($\mathbf{x}$) represents an $N$-dimensional vector with elements $\frac{x_1}{\left | x_1 \right |},\cdots ,\frac{x_N}{\left | x_N \right |}$. $\mathbf{A}\circ\mathbf{B}$ represents the Hadamard product of matrices $\mathbf{A}$ and $\mathbf{B}$. $\Re\left \{ \cdot \right \}$ extracts the real value of a complex variable. $\mathcal{E}\left \{ \cdot \right \}$ denotes statistical expectation. $\overset{\Delta }{=}$ and $\sim$ stand for ``defined as'' and ``distributed as'', respectively. The distribution of a circularly symmetric complex Gaussian random variable with mean $\mu$ and variance $\sigma^2$ is denoted by $\mathcal{CN}(\mu ,\sigma^2)$. $[x]^+$ stands for $\mathrm{max}\left \{ 0,x \right \}$. The gradient vector of function $f(\mathbf{x})$ with respect to $\mathbf{x}$ is denoted by $\nabla_{\mathbf{x}} f(\mathbf{x})$.
\subsection{IRS-assisted Multiuser  Wireless Communication System}
\begin{figure}[t]\vspace*{-2mm}
\centering
\includegraphics[width=2.3in]{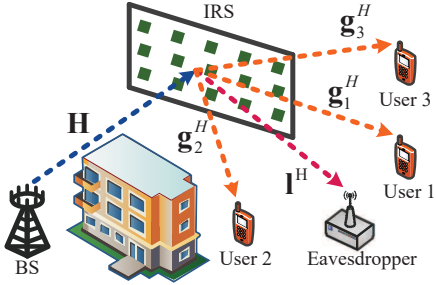} \vspace*{-3mm}
\caption{An intelligent reflecting surface (IRS)-assisted secure communication system with one eavesdropper and $K=3$ desired users. The direct links from the BS to the users and the eavesdropper are blocked by a building.}
\label{fig:IRS-model}\vspace*{-5mm}
\end{figure}\vspace*{-0mm}
We consider an IRS-assisted multiuser DL communication system which comprises a BS, an eavesdropper, an IRS, and a set of desired users, indexed by $\mathcal{K}\overset{\Delta }{=}\left \{1,\cdots ,K \right \}$, as illustrated in Figure \ref{fig:IRS-model}. The BS is equipped with $N_{\mathrm{T}}>1$ antennas, while both the desired users and the eavesdropper are single-antenna devices. Moreover, a passive IRS is deployed to achieve secure communication between the BS and the users. The IRS employs $M$ phase shifters, indexed by $\mathcal{M}\overset{\Delta }{=}\left \{1,\cdots ,M \right \}$, and can be programmed and reconfigured via a controller. Furthermore, perfect channel state information (CSI) of the whole system is assumed to be available at the BS for resource allocation design\footnote{In practice, the BS may not be able to obtain perfect CSI. Hence, the results in this paper serve as a theoretical system performance benchmark.}. Besides, we assume that the direct links from the BS to the users and the eavesdropper are unavailable due to unfavorable propagation conditions (e.g., blockage by a building). 
\par
In each scheduling time slot, the BS transmits a signal vector $\mathbf{x}\in \mathbb{C}^{\mathit{N}_{\mathrm{T}}}$ to the $K$ users. In particular, the signal vector, which comprises $K$ information signals and AN, is given by
\vspace*{-2mm}
\begin{equation}
\mathbf{x}=\underset{k\in\mathcal{K}}{\sum }\mathbf{w}_ks_k+\mathbf{z},\\[-2mm]
\end{equation}
where $\mathbf{w}_k\in \mathbb{C}^{\mathit{N}_{\mathrm{T}}}$ and ${s}_k\in\mathbb{C}$ denote the beamforming vector for the $k$-th user and the corresponding information bearing signal, respectively. We assume $\mathcal{E}\{\left |s _{k} \right|^2\}=1$, $\forall\mathit{k} \in \mathcal{K}$, without loss of generality. Moreover, to guarantee secure communication, an AN vector $\mathbf{z}\in \mathbb{C}^{\mathit{N}_{\mathrm{T}}}$ is generated and transmitted by the BS to impair the eavesdropper. In
particular, we model $\mathbf{z}$ as a complex Gaussian random vector with zero mean and covariance matrix $\mathbf{Z}\in\mathbb{H}^{\mathit{N}_{\mathrm{T}}}$, $\mathbf{Z}\succeq \mathbf{0}$. 
\par
The signals received by user $k$ and the
eavesdropper via the reflection at the IRS are given by
\vspace*{-2mm}
\begin{eqnarray}
y_k&=&\mathbf{g}_k^H\mathbf{\Phi}\mathbf{H}(\underset{k\in\mathcal{K}}{\sum }\mathbf{w}_ks_k+\mathbf{z})+n_k,\\
y_e&=&\mathbf{l}^H\mathbf{\Phi}\mathbf{H}(\underset{k\in\mathcal{K}}{\sum }\mathbf{w}_ks_k+\mathbf{z})+n_e,\\[-4mm]\notag
\end{eqnarray}
respectively, where $\mathbf{g}_k\in \mathbb{C}^M$ and $\mathbf{l}\in \mathbb{C}^M$ denote the channel vectors between the IRS and user $k$ and between the IRS and the eavesdropper, respectively. $\mathbf{\Phi}=\mathrm{diag}\left ( e^{j\phi_1}, \cdots, e^{j\phi_M}  \right )$ denotes the phase shift matrix of the IRS, where $\phi_m$, $\forall m \in \mathcal{M}$, represents the phase shift of the $m$-th reflector of the IRS \cite{wu2018intelligent}. The channel matrix between the BS and the IRS is denoted by $\mathbf{H}\in\mathbb{C}^{{\mathit{M}\times\mathit{N}_{\mathrm{T}}}}$. Besides, $n_k\sim \mathcal{CN}(0, \sigma_{n_k}^2)$ and $n_e\sim \mathcal{CN}(0, \sigma_{n_e}^2)$ are the additive white Gaussian noise samples at user $k$ and the eavesdropper, respectively.

\section{Optimization Problem Formulation}
In this section, we first define the adopted system performance metric and then formulate the resource allocation optimization problem for the considered system. 

\subsection{Achievable Rate and Secrecy Rate}
The achievable rate (bits/s/Hz) of user $k$ is given by $R_k=\mathrm{log}_2(1+\Gamma_k)$, where
\vspace*{-2mm}
\begin{equation}
\Gamma_k\hspace*{-0.5mm}=\hspace*{-0.5mm}\frac{\left |\mathbf{g}_k^H\mathbf{\Phi}\mathbf{H}\mathbf{w}_k\right |^2}{\hspace*{-3mm}\underset{r\in\mathcal{K}\setminus \left \{ k \right \}}{\sum }\hspace*{-3mm}\left |\mathbf{g}_k^H\mathbf{\Phi}\mathbf{H}\mathbf{w}_r\right |^2\hspace*{-0.5mm}+\hspace*{-0.5mm}\mathrm{Tr}(\mathbf{H}^H\mathbf{\Phi}^H\mathbf{g}_k\mathbf{g}_k^H\mathbf{\Phi}\mathbf{H}\mathbf{Z})\hspace*{-1mm}+\hspace*{-1mm} \sigma^2_{n_k}}.\\[-1mm]
\end{equation}
In this paper, we impose a worst-case assumption regarding the capabilities of the eavesdropper for resource allocation algorithm design to ensure secure communication \cite{7955970}. Specifically, we assume that the eavesdropper is capable of canceling all multiuser interference before decoding the desired information. Therefore, the channel capacity between the BS and the eavesdropper for wiretapping user $k$ is given by
\vspace*{-2mm}
\begin{equation}
\label{evec1}
    C_k^{\mathrm{E}}=\mathrm{log}_2\Big(1+\frac{\left |\mathbf{l}^H\mathbf{\Phi}\mathbf{H}\mathbf{w}_k\right |^2}{{\mathrm{Tr}(\mathbf{H}^H\mathbf{\Phi}^H\mathbf{l}\mathbf{l}^H\mathbf{\Phi}\mathbf{H}\mathbf{Z})}+  \sigma^2_{n_e}}\Big).\\[-1mm]
\end{equation}
The achievable secrecy rate between the BS and user $k$ is given by $R_k^{\mathrm{Sec}}=\left [R_k-C_k^{\mathrm{E}} \right ]^+$ \cite{7463025}.
\subsection{Optimization Problem Formulation}
We aim to maximize the system sum secrecy rate by optimizing $\mathbf{w}_k$, $\mathbf{Z}$, and $\mathbf{\Phi}$. The corresponding optimization problem is formulated as
\vspace*{-1mm}
\begin{eqnarray}
\label{prob1}
&&\underset{\mathbf{w}_{\mathit{k}},\mathbf{Z}\in\mathbb{H}^{\mathit{N}_{\mathrm{T}}},\mathbf{\Phi}}{\maxo} \,\, \,\, \underset{ k\in\mathcal{K}}{\sum} \left [R_k-C_k^{\mathrm{E}} \right ]^+ \\
\mbox{s.t.}\hspace*{-4mm}
&&\mbox{C1:}\underset{k\in\mathcal{K}}{\sum }\left \| \mathbf{w}_k \right \|^2+\mathrm{Tr}(\mathbf{Z})\leq P_{\mathrm{max}},\notag\\
&&\mbox{C2: }
\Big |[\mathbf{\Phi}]_{m,m}\Big |=1, \forall m,~~\mbox{C3: }\mathbf{Z}\succeq \mathbf{0}.\notag\\[-6mm]\notag
\end{eqnarray}
Constraint C1 limits the maximum BS transmit power allowance to $P_{\mathrm{max}}$. 
Besides, the operator $\left [ \cdot  \right ]^+$ has no impact on the optimal solution and hence is omitted in the following for notational simplicity\footnote{It can be proved that at the optimal solution, if the achievable secrecy rate of user $k$ is non-positive, the proposed algorithm would turn off the transmission of user $k$ and reallocate the available power to other users.}.
\par
We note that it is very arduous to obtain the globally optimal solution of \eqref{prob1}, due to the coupling of the optimization variables and the unit modulus constraint in C2. Therefore, we develop a resource allocation algorithm based on alternating optimization \cite{bezdek2002some} to obtain a suboptimal solution of \eqref{prob1} in the next section.

\section{Solution of the Problem}
In this section, we aim to design a computationally efficient suboptimal algorithm for handling \eqref{prob1} via alternating optimization. For notational simplicity, we first define $\mathbf{G}_k=\mathrm{diag}(\mathbf{g}_k^H)\mathbf{H}$, $\mathbf{L}=\mathrm{diag}(\mathbf{l}^H)\mathbf{H}$, $\mathbf{W}_k=\mathbf{w}_k\mathbf{w}_k^H$. Moreover, we define a new optimization variable $\mathbf{u}=\left [ e^{j\phi_1}, \cdots, e^{j\phi_M}  \right ]^H$. Then, we rewrite the received SINRs at user $k$ as follows:
\vspace*{-1mm}
\begin{equation}
\hspace*{-2mm}\Gamma_k\hspace*{-0.5mm}=\hspace*{-0.5mm}\frac{\mathrm{Tr}(\mathbf{W}_k\mathbf{G}_k^H\mathbf{u}\mathbf{u}^H\mathbf{G}_k)}{\hspace*{-3.5mm}\underset{r\in\mathcal{K}\setminus \left \{ k \right \}}{\sum }\hspace*{-3.5mm}\mathrm{Tr}(\mathbf{W}_r\mathbf{G}_k^H\mathbf{u}\mathbf{u}^H\mathbf{G}_k)\hspace*{-0.8mm}+\hspace*{-0.8mm}\mathrm{Tr}(\mathbf{Z}\mathbf{G}_k^H\mathbf{u}\mathbf{u}^H\mathbf{G}_k)\hspace*{-0.8mm}+\hspace*{-0.8mm}  \sigma^2_{n_k}}.\\[-1mm] 
\end{equation}
Moreover, the channel capacity for the eavesdropper with respect to the message of user $k$ in (\ref{evec1}) can be rewritten as
\vspace*{-1mm}
\begin{equation}
    \label{evec2}
    C_k^{\mathrm{E}}=\mathrm{log}_2\Big(1+\frac{\mathrm{Tr}(\mathbf{W}_k\mathbf{L}^H\mathbf{u}\mathbf{u}^H\mathbf{L})}{\mathrm{Tr}(\mathbf{Z}\mathbf{L}^H\mathbf{u}\mathbf{u}^H\mathbf{L})+  \sigma^2_{n_e}}\Big).\\[-1mm]
\end{equation}
Now, to facilitate the application of alternating optimization, we first recast \eqref{prob1} in equivalent form as follows:
\vspace*{-1mm}
\begin{eqnarray}
\label{prob2}
&&\underset{\mathbf{Z}\in\mathbb{H}^{\mathit{N}_{\mathrm{T}}},\mathbf{W},\mathbf{u}}{\mino} \,\, \,\, f=F_1+F_2-G_1-G_2 \\
\notag\mbox{s.t.}\hspace*{-4mm}
&&\mbox{C1:}\underset{k\in\mathcal{K}}{\sum }\mathrm{Tr}(\mathbf{W}_k)+\mathrm{Tr}(\mathbf{Z})\leq P_{\mathrm{max}},\notag\\
&&\mbox{C2: }
\left | u_m \right |=1,~\forall m,\hspace*{6mm}\mbox{C3: }\mathbf{Z}\succeq \mathbf{0},\notag\\
&&\mbox{C4: }\mathbf{W}_k\succeq \mathbf{0},~\forall k,\hspace*{8.3mm}\mbox{C5: }\mathrm{Rank}(\mathbf{W}_k)\leq 1,~\forall k,\notag\\[-7mm]\notag
\end{eqnarray}
where $\mathbf{W}\in\mathbb{C}^{K\times N_{\mathrm{T}}}$ are the collection of all $\mathbf{W}_k$, and $F_1$, $F_2$, $G_1$, and $G_2$ are shown at the bottom of this page. Moreover, $u_m$ is the $m$-th element of $\mathbf{u}$, and $\mathbf{W}_{\mathit{k}}\in\mathbb{H}^{\mathit{N}_{\mathrm{T}}}$, $\mathbf{W}_k\succeq \mathbf{0}$, and $\mathrm{Rank}(\mathbf{W}_k)\leq 1$ in (\ref{prob2}) are imposed to ensure that $\mathbf{W}_k=\mathbf{w}_k\mathbf{w}_k^H$ holds after optimization. 
\par
By employing alternating optimization, we iteratively
optimize $\left \{ \mathbf{W},\mathbf{Z}\right \}$ and $\mathbf{u}$ with the other one fixed. In particular, for a given $\mathbf{u}$, we solve \eqref{prob2} by employing successive convex approximation (SCA) \cite{dinh2010local} and semidefinite relaxation (SDR) \cite{7463025}. Then, for given $\mathbf{W}$ and $\mathbf{Z}$, we solve for $\mathbf{u}$ by applying manifold optimization \cite{absil2009optimization}.
\begin{figure*}[b]
	\setcounter{TempEqCnt}{\value{equation}} 		 
	\setcounter{equation}{\value{equation}}
	\hrule
\begin{eqnarray}
\hspace*{-6mm}F_1&&\hspace*{-7mm}=\hspace*{-0.5mm}-\hspace*{-1mm}\underset{k\in\mathcal{K}}{\hspace*{-0.3mm}\sum }\mathrm{log}_2\Big(\underset{r\in\mathcal{K}}{\sum }\mathrm{Tr}(\mathbf{W}_r\mathbf{G}_k^H\mathbf{u}\mathbf{u}^H\mathbf{G}_k)\hspace*{-0.6mm}+\hspace*{-0.6mm}\mathrm{Tr}(\mathbf{Z}\mathbf{G}_k^H\mathbf{u}\mathbf{u}^H\mathbf{G}_k)\hspace*{-0.6mm}+\hspace*{-0.6mm}\sigma^2_{n_k}\Big),
F_2=\hspace*{-1mm}-K\mathrm{log}_2\Big(\mathrm{Tr}(\mathbf{Z}\mathbf{L}^H\mathbf{u}\mathbf{u}^H\mathbf{L})+ \sigma^2_{n_e}\Big),\\
\hspace*{-6mm}G_1&&\hspace*{-7mm}=\hspace*{-1mm}-\hspace*{-1mm}\underset{k\in\mathcal{K}}{\sum }\mathrm{log}_2\Big(\hspace*{-3mm}\underset{r\in\mathcal{K}\setminus\left \{ k \right \}}{\sum }\hspace*{-3mm}\mathrm{Tr}(\mathbf{W}_r\mathbf{G}_k^H\mathbf{u}\mathbf{u}^H\hspace*{-0.3mm}\mathbf{G}_k)\hspace*{-1mm}+\hspace*{-1mm}\mathrm{Tr}(\mathbf{Z}\mathbf{G}_k^H\mathbf{u}\mathbf{u}^H\mathbf{G}_k)\hspace*{-1mm}+\hspace*{-1mm}\sigma^2_{n_k}\hspace*{-1mm}\Big),
G_2\hspace*{-1mm}=\hspace*{-1mm}-\hspace*{-0.8mm}\underset{k\in\mathcal{K}}{\sum }\mathrm{log}_2\Big(\hspace*{-0.3mm}\mathrm{Tr}(\mathbf{W}_k\mathbf{L}^H\hspace*{-0.3mm}\mathbf{u}\mathbf{u}^H\hspace*{-0.3mm}\mathbf{L})\hspace*{-1mm}+\hspace*{-1mm}{\mathrm{Tr}(\mathbf{Z}\mathbf{L}^H\hspace*{-0.3mm}\mathbf{u}\mathbf{u}^H\hspace*{-0.3mm}\mathbf{L})\hspace*{-1mm}+ \hspace*{-1mm}\sigma^2_{n_e}}\hspace*{-1mm}\Big)\hspace*{-0.3mm}
\end{eqnarray}\vspace*{-5mm}
\end{figure*}
\setcounter{equation}{\value{TempEqCnt}+2}
\subsection{SCA and SDR}
For a given $\mathbf{u}$, the optimization problem in (\ref{prob2}) can be rewritten as
\vspace*{-5mm}
\begin{eqnarray}
\label{prob3}
&&\hspace*{-8mm}\underset{\mathbf{Z}\in\mathbb{H}^{\mathit{N}_{\mathrm{T}}},\mathbf{W}}{\mino} \,\, \,\, F_1+F_2-G_1-G_2 \\
&&\hspace*{-2mm}\mbox{s.t.}\hspace*{7mm}\mbox{C1},\mbox{C3-C5}.\notag\\[-7mm]\notag
\end{eqnarray}
To facilitate the application of SCA, we first construct global underestimators of $G_1$ and $G_2$, respectively \cite{dinh2010local}. In particular, for any feasible point $\mathbf{W}^i$ and $\mathbf{Z}^i$, the differentiable convex function $G_1(\mathbf{W},\mathbf{Z})$ satisfies the following inequality:
\vspace*{-2mm}
\begin{eqnarray}
\label{G1}
G_1(\mathbf{W},\mathbf{Z})&&\hspace*{-6mm}\geq G_1(\mathbf{W}^i,\mathbf{Z}^i)\notag\\&&\hspace*{-6mm}+\mathrm{Tr}\Big(\big(\nabla_{\mathbf{W}}G_1(\mathbf{W}^i,\mathbf{Z}^i)\big)^H(\mathbf{W}-\mathbf{W}^i)\Big)\notag\\&&\hspace*{-6mm}+\mathrm{Tr}\Big(\big(\nabla_{\mathbf{Z}}G_1(\mathbf{W}^i,\mathbf{Z}^i)\big)^H(\mathbf{Z}-\mathbf{Z}^i)\Big)\notag\\&&\hspace*{-6mm}\overset{\Delta }{=}\widetilde{G_1}(\mathbf{W},\mathbf{Z},\mathbf{W}^i,\mathbf{Z}^i),\\[-7mm]\notag
\end{eqnarray}
where the right hand side term in (\ref{G1}) is a global underestimation of $G_1(\mathbf{W},\mathbf{Z})$. Similarly, a global underestimation of $G_2(\mathbf{W},\mathbf{Z})$ at feasible point $\mathbf{W}^i$ and $\mathbf{Z}^i$ can be constructed as follows
\vspace*{-2mm}
\begin{eqnarray}
\label{G2}
\widetilde{G_2}(\mathbf{W},\mathbf{Z},\mathbf{W}^i,\mathbf{Z}^i)&&\hspace*{-6mm}\overset{\Delta }{=}G_2(\mathbf{W}^i,\mathbf{Z}^i)\notag\\&&\hspace*{-6mm}+\mathrm{Tr}\Big(\big(\nabla_{\mathbf{W}}G_2(\mathbf{W}^i,\mathbf{Z}^i)\big)^H(\mathbf{W}-\mathbf{W}^i)\Big)\notag\\&&\hspace*{-6mm}+\mathrm{Tr}\Big(\big(\nabla_{\mathbf{Z}}G_2(\mathbf{W}^i,\mathbf{Z}^i)\big)^H(\mathbf{Z}-\mathbf{Z}^i)\Big).\\[-6mm]\notag
\end{eqnarray}
Therefore, for any given $\mathbf{W}^i$ and $\mathbf{Z}^i$, an upper bound of (\ref{prob3}) can be obtained by solving the following optimization problem:
\vspace*{-5mm}
\begin{eqnarray}
\label{prob4}
&&\hspace*{-8mm}\underset{\mathbf{Z}\in\mathbb{H}^{\mathit{N}_{\mathrm{T}}},\mathbf{W}}{\mino} \,\, \,\, F_1+F_2-\widetilde{G_1}-\widetilde{G_2} \\
&&\hspace*{-2mm}\mbox{s.t.}\hspace*{7mm}\mbox{C1},\mbox{C3-C5}.\notag\\[-7mm]\notag
\end{eqnarray}
We note that the remaining non-convexity of \eqref{prob4} stems from the rank-one constraint $\mbox{C5}$. To tackle this issue, we remove constraint $\mbox{C5}$ by applying SDR where the relaxed version of \eqref{prob4} can be efficiently solved via convex problem solvers such as CVX \cite{grant2008cvx}. In the following theorem, we reveal the tightness of SDR.
\par
\textit{Theorem 1:~}If $\mathrm{P}_{\mathrm{max}}>0$, an optimal beamforming matrix $\mathbf{W}_k$ satisfying $\mathrm{Rank}(\mathbf{W}_k)\leq 1$ can always be obtained.
\par
\textit{Proof:~}Please refer to the Appendix. \qed
\par
We note that the minimum of \eqref{prob4} serves as an upper bound of \eqref{prob3}. By employing the algorithm summarized in \textbf{Algorithm 1}, we can iteratively tighten the upper bound and obtain a sequence of solutions $\mathbf{W}$ and $\mathbf{Z}$. It can be shown that the objective function in \eqref{prob4} is non-increasing in each iteration, and the developed algorithm is guaranteed to converge to a locally optimal solution of \eqref{prob3} \cite{dinh2010local}.
\vspace*{2mm}
\begin{algorithm}[t]
\caption{Successive Convex Approximation-Based Algorithm}
\begin{algorithmic}[1]
\small
\STATE Initialize iteration index $i=1$.
\REPEAT
\STATE Solve \eqref{prob4} for given $\mathbf{W}^i$ and $\mathbf{Z}^i$ and store the intermediate solution ${\mathbf{W},\mathbf{Z}}$

\STATE Set $i=i+1$ and $\mathbf{W}^i=\mathbf{W}$ and $\mathbf{Z}^i=\mathbf{Z}$
\UNTIL convergence
\STATE $\mathbf{W}^*=\mathbf{W}^i$ and $\mathbf{Z}^*=\mathbf{Z}^i$
\end{algorithmic}
\end{algorithm}
\vspace*{2mm}
\subsection{Oblique Manifold Optimization}
\begin{figure*}\vspace*{3mm}
	\centering
	\subfigure[Tangent space and Riemannian gradient.]
	{
		\centering\includegraphics[height=4cm]{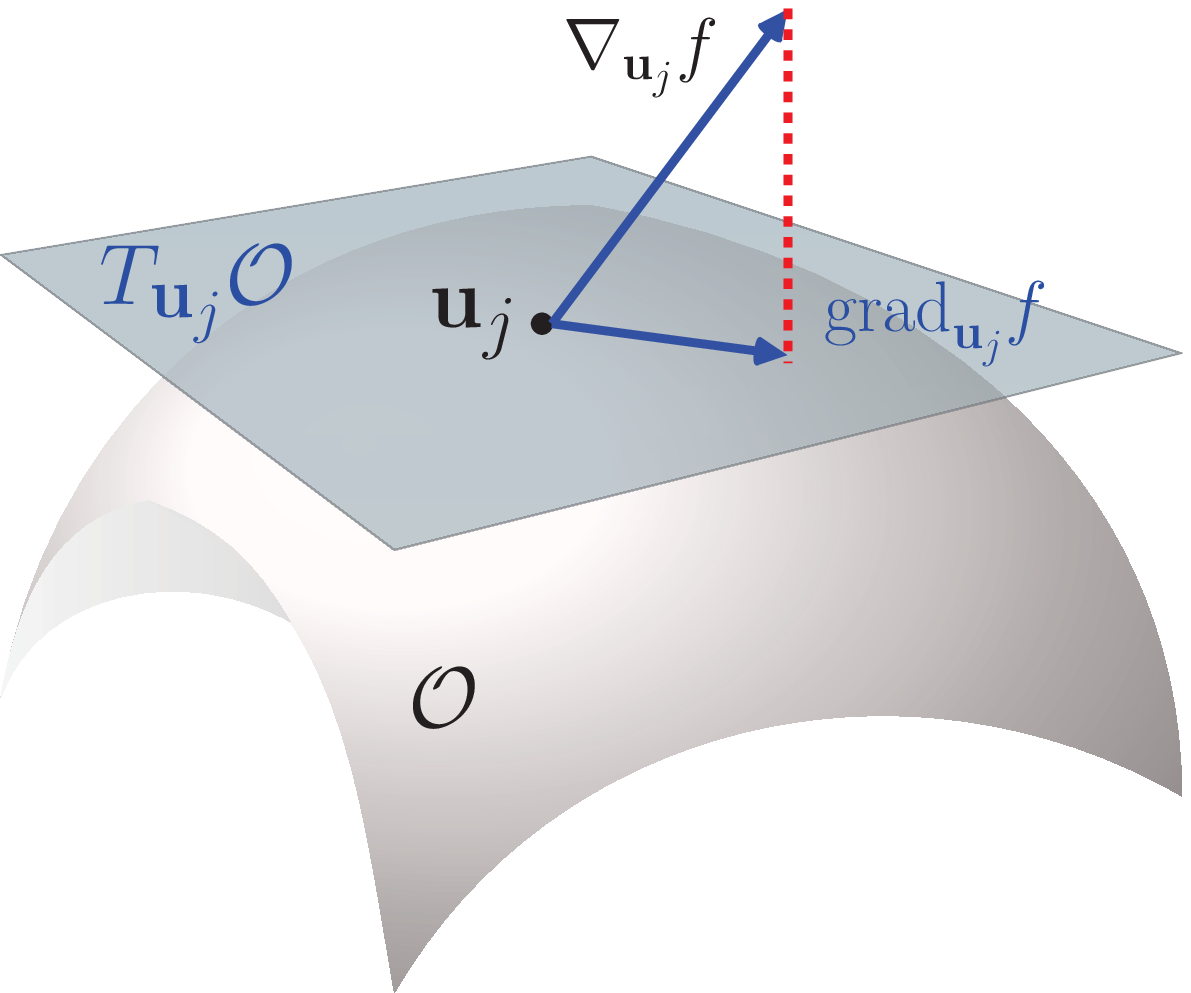}\label{fig21}
	}\quad\quad
	\subfigure[Vector transport.]
	{
		\centering\includegraphics[height=4cm]{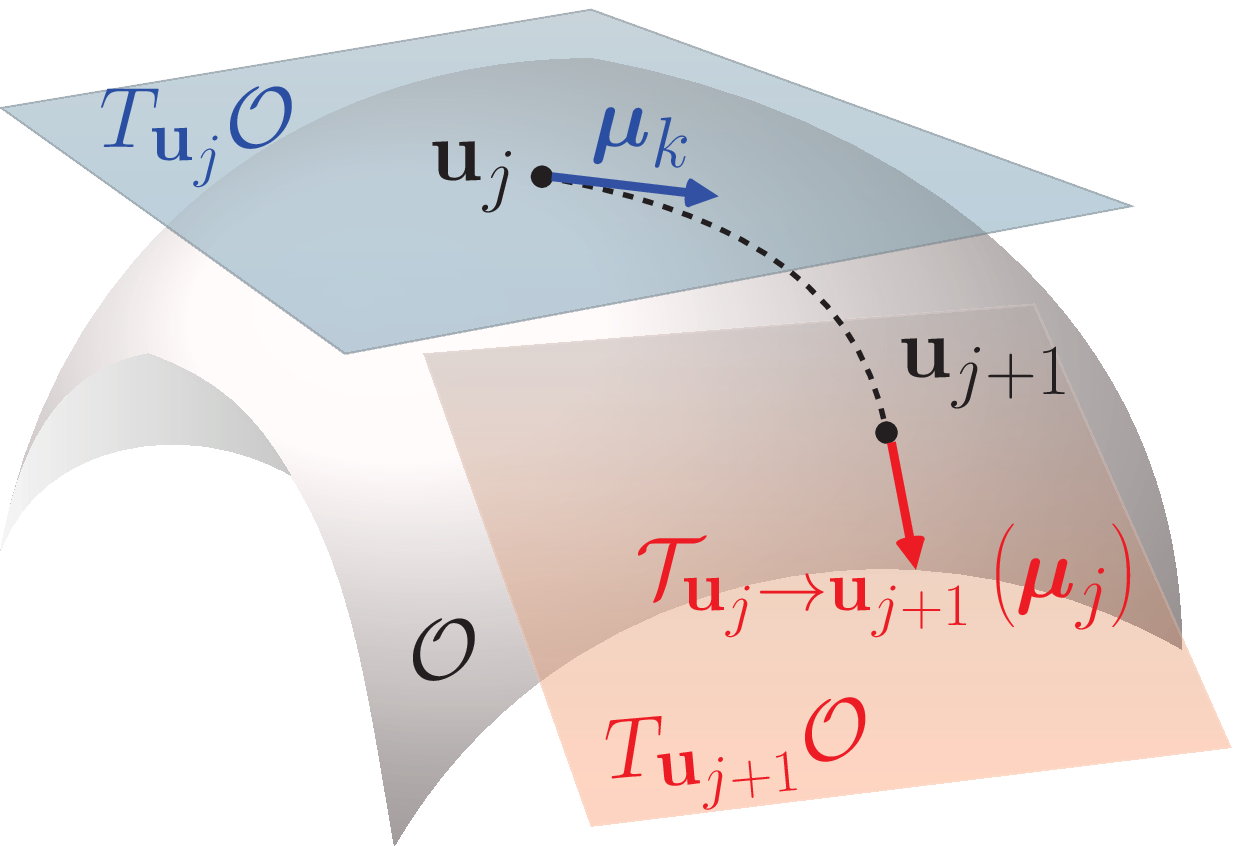}\label{fig22}
	}\quad\quad
		\subfigure[Retraction.]
	{
		\centering\includegraphics[height=4cm]{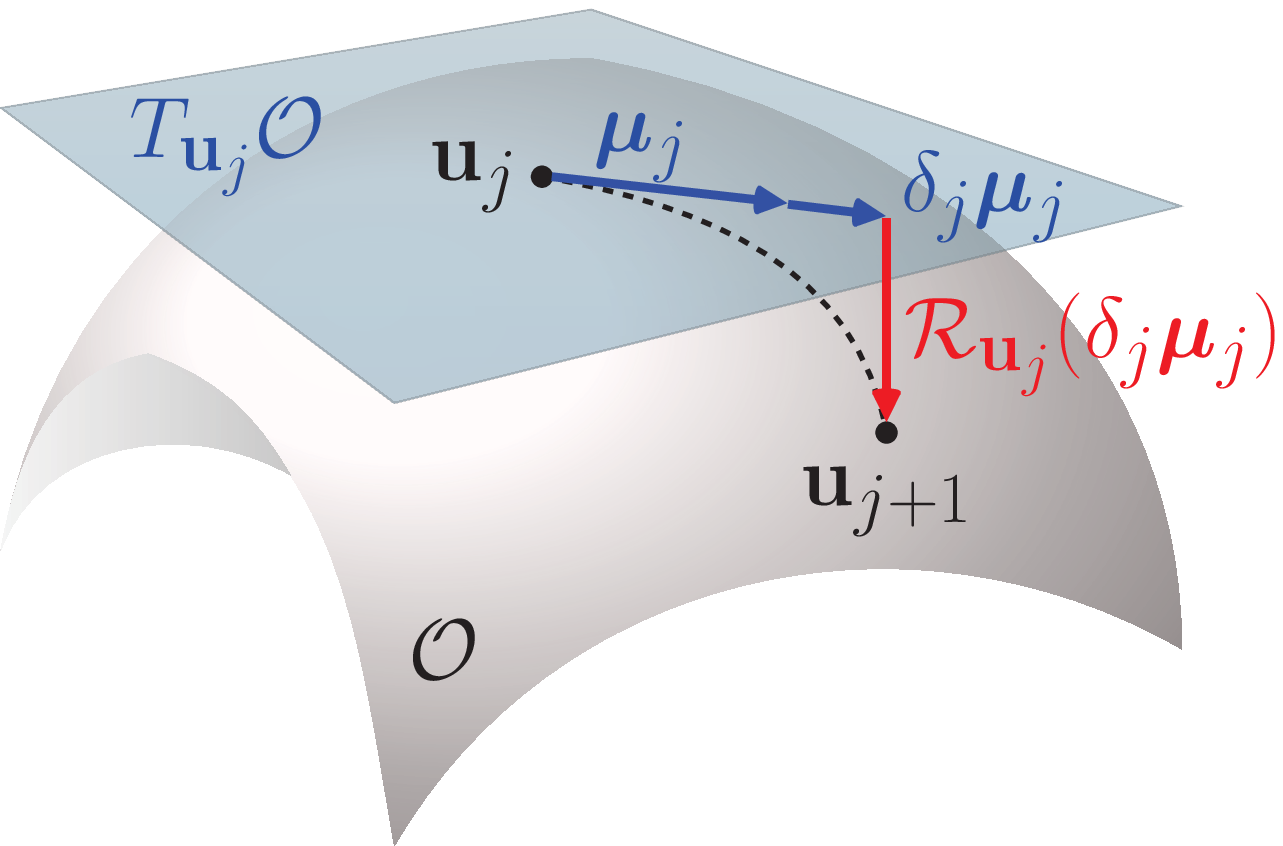}\label{fig23}
	}\vspace*{-2mm}
	\caption{An illustration of major definitions in Riemannian manifold optimization.}\label{manifold}
\vspace*{-3mm}
\end{figure*}
For given $\mathbf{W}_k$ and $\mathbf{Z}$, we can rewrite \eqref{prob2} as:
\vspace*{-2mm}
\begin{eqnarray}
\label{prob5}
&&\hspace*{-8mm}\underset{\mathbf{u}}{\mino} \,\,\,\,
F_1+F_2-G_1-G_2 \\
&&\hspace*{-2mm}\mbox{s.t.}\hspace*{7mm}\mbox{C2: }[\mathbf{u}\mathbf{u}^H]_{m,m}=1,~\forall m.\notag\\[-7mm]\notag
\end{eqnarray}
We note that it is very challenging to solve \eqref{prob5} optimally due to the non-convex unit modulus constraint C2. In the literature, the unit modulus constraint is often handled by SDR and Gaussian randomization \cite{wu2018intelligent} which leads to a suboptimal solution. Yet, the objective function may not be monotonically non-increasing in each iteration when this approach is applied. Thus, the corresponding algorithm cannot guarantee convergence.
In contrast, in this paper, we develop a manifold optimization-based algorithm which is guaranteed to converge to a suboptimal solution. Moreover, unlike \cite{chen2019intelligent} where the unit modulus constraint was relaxed, in this paper, the unit modulus constraint is handled directly by exploiting manifold optimization theory \cite{absil2009optimization}. We note that constraint C2 defines an oblique manifold \cite{absil2009optimization} which can be characterized by
\vspace*{-1mm}
\begin{equation}
    \mathcal{O}=\left \{\mathbf{u}\in\mathbb{C}^M~|~ [\mathbf{u}\mathbf{u}^H]_{m,m}=1,~\forall m\in\mathcal{M}\right \}.\\[-1mm]
\end{equation}
We note that constraint C2 is automatically satisfied when optimizing $\mathbf{u}$ over the oblique manifold. Now, we introduce some definitions which are commonly used in Riemannian manifold optimization \cite{absil2009optimization}.
\par
The \textit{tangent space} of the oblique manifold $\mathcal{O}$ at point $\mathbf{u}_j$ is defined as the space which contains all tangent vectors of the oblique manifold $\mathcal{O}$ at point $\mathbf{u}_j$, cf. Figure \ref{fig21}. Specifically, each tangent vector is a vector that is a tangent to the oblique manifold $\mathcal{O}$ at point $\mathbf{u}_j$ \cite{absil2009optimization}. The tangent space for $\mathcal{O}$ at $\mathbf{u}_j$ is given by
\vspace*{-1mm}
\begin{equation}
    T_{\mathbf{u}_j}\mathcal{O}=\left \{\mathbf{v}\in\mathbb{C}^M~ |~ [\mathbf{v}\mathbf{u}_j^H]_{m,m}=0,~\forall m\in\mathcal{M}\right \},\\[-1mm]
\end{equation}
where $\mathbf{v}$ is a tangent vector at $\mathbf{u}_j$. Among all tangent vectors, the one that yields the fastest increase of the objective function is defined as the \textit{Riemannian gradient}, i.e., $\mathrm{grad}_{\mathbf{u}_j}f$. The Riemannian gradient of function $f$ at point $\mathbf{u}_j$ is calculated based on
the orthogonal projection of the Euclidean gradient $\nabla_{\mathbf{u}_j}f$ onto tangent space $T_{\mathbf{u}_j}\mathcal{O}$ \cite{gallot1990riemannian}. In particular, $\mathrm{grad}_{\mathbf{u}_j}f$ is given by
\vspace*{-1mm}
\begin{equation}
\label{obmgrad}
    \mathrm{grad}_{\mathbf{u}_j}f=\nabla_{\mathbf{u}_j}f-\Re \left \{ \nabla_{\mathbf{u}_j}f\circ (\mathbf{u}_j^T)^H \right \}\circ \mathbf{u}_j,\\[-1mm]
\end{equation}
where $\nabla_{\mathbf{u}_j}f$ is obtained as \cite{boyd2004convex}
\vspace*{-2mm}
\begin{eqnarray}
\nabla_{\mathbf{u}_j}f&&\hspace{-6mm}=\frac{K\mathbf{L}(\mathbf{Z}^H+\mathbf{Z})\mathbf{L}^H\mathbf{u}_j}{(\mathrm{ln}2)F_2(\mathbf{u}_j)}\notag\\&&\hspace{-6mm}+\frac{\underset{ k\in\mathcal{K}}{\sum}\hspace*{1mm}\underset{ r\in\mathcal{K}}{\sum}\big[\mathbf{G}_k(\mathbf{W}_r^H+\mathbf{W}_r+\mathbf{Z}^H+\mathbf{Z})\mathbf{G}_k^H\mathbf{u}_j\big]}{(\mathrm{ln}2)F_1(\mathbf{u}_j)}\notag\\&&\hspace{-6mm}-\frac{\underset{ k\in\mathcal{K}}{\sum}\hspace*{1mm}\underset{ r\in\mathcal{K}\setminus \left \{k\right \}}{\sum}\big[\mathbf{G}_k(\mathbf{W}_r^H+\mathbf{W}_r+\mathbf{Z}^H+\mathbf{Z})\mathbf{G}_k^H\mathbf{u}_j\big]}{(\mathrm{ln}2)G_1(\mathbf{u}_j)}\notag\\&&\hspace*{-6mm}-\frac{\underset{ k\in\mathcal{K}}{\sum}\mathbf{L}(\mathbf{W}_k^H+\mathbf{W}_k+\mathbf{Z}^H+\mathbf{Z})\mathbf{L}^H\mathbf{u}_j}{(\mathrm{ln}2)G_2(\mathbf{u}_j)}. \\[-5mm]\notag
\end{eqnarray}
After obtaining the Riemannian gradient $\mathrm{grad}_{\mathbf{u}_j}f$, we can exploit the
optimization approaches designed for the Euclidean space
to tackle manifold optimization problems. In particular, we employ the conjugate gradient (CG) method \cite{avriel2003nonlinear}, where the update rule of the search direction in the Euclidean space is given by
\vspace*{-2mm}
\begin{equation}
\label{euupdate}
\bm{\mu}_{j+1}=-\nabla_{\mathbf{u}_{j+1}}f+\alpha_j\bm{\mu}_{j}.\\[-1mm]
\end{equation}
Here, $\bm{\mu}_{j}$
denotes the search direction at $\mathbf{u}_j$ and $\alpha_j$ is chosen as the Polak-Ribi\`ere parameter to achieve fast convergence \cite{avriel2003nonlinear}. However, since vectors $\bm{\mu}_{j}$ and $\bm{\mu}_{j+1}$ in \eqref{euupdate} lie in $T_{\mathbf{u}_j}\mathcal{O}$ and $T_{\mathbf{u}_{j+1}}\mathcal{O}$, respectively, they cannot be integrated directly over different tangent spaces. To circumvent this problem, we introduce an operation called \textit{transport} which
maps $\bm{\mu}_{j}$ from tangent space $T_{\mathbf{u}_j}\mathcal{O}$ to tangent space $T_{\mathbf{u}_{j+1}}\mathcal{O}$ \cite{7397861}. In particular, the vector transport for oblique manifold $\mathcal{O}$, as shown in cf. Figure \ref{fig22}, is given by
\vspace*{-3mm}
\begin{eqnarray}
\label{transport}
\hspace*{-4mm}\mathcal{T}_{\mathbf{u}_j\rightarrow \mathbf{u}_{j+1}}(\bm{\mu}_{j})
\overset{\Delta}{=}T_{\mathbf{u}_j}\mathcal{O}\hspace*{-4mm}&&\hspace*{-2mm}\mapsto T_{\mathbf{u}_{j+1}}\mathcal{O}:\notag\\
\bm{\mu}_j \hspace*{-4mm}&&\hspace*{-2mm}\mapsto \bm{\mu}_{j}\hspace*{-1mm}-\hspace*{-1mm}\Re \left \{\bm{\mu}_{j}\hspace*{-0.5mm}\circ\hspace*{-0.5mm}(\mathbf{u}^T_{j+1})^H\right \}\hspace*{-0.5mm}\circ\hspace*{-0.5mm} \mathbf{u}_{j+1}.\\[-6mm]\notag
\end{eqnarray}
Similar to \eqref{euupdate}, the search direction of the Riemannian gradient in \eqref{obmgrad} can be updated based on the following equation:
\begin{equation}
\label{obmupdate}
\bm{\mu}_{j+1}=-\mathrm{grad}_{\mathbf{u}_{j+1}}f+\alpha_j\mathcal{T}_{\mathbf{u}_j\rightarrow \mathbf{u}_{j+1}}(\bm{\mu}_{j}).\\[-2mm]
\end{equation}
After determining the search direction $\bm{\mu}_{j}$
at $\mathbf{u}_{j}$, we introduce another operation called \textit{retraction} to determine the destination on the oblique manifold \cite{7397861}. In other words, by applying retraction, we
map a vector in the tangent space $T_{\mathbf{u}_j}\mathcal{O}$ onto the manifold
$\mathcal{O}$, cf. Figure \ref{fig23}. In particular, for a given point $\mathbf{u}_{j}$ on manifold $\mathcal{O}$, the retraction for step size $\delta_j$ 
and search direction $\bm{\mu}_{j}$ are given as
\vspace*{-2mm}
\begin{equation}
\label{retraction}
 \mathcal{R}_{\mathbf{u}_{j}}(\delta_j\bm{\mu}_{j})\overset{\Delta }{=}T_{\mathbf{u}_j}\mathcal{O}\mapsto\mathcal{O}:\delta_j\bm{\mu}_{j}\mapsto\mathrm{unt}(\delta_j\bm{\mu}_{j}).\\[-2mm]
\end{equation}
The problem in \eqref{prob5} can be tackled by applying the proposed algorithm summarized in \textbf{Algorithm 2}. Since \textbf{Algorithm 2} is a gradient-based algorithm, the objective function in \eqref{prob5} is monotonically non-increasing in each iteration. Hence, \textbf{Algorithm 2} is guaranteed to converge to a stationary point of \eqref{prob5} \cite{avriel2003nonlinear}.  
\begin{algorithm}[t]
\caption{Oblique Manifold Optimization-Based Algorithm}
\begin{algorithmic}[1]
\small
\STATE Set iteration index $j=1$, convergence tolerance $\varepsilon$, step size $\delta_j$, and initial point $\mathbf{u}_1$
\STATE Calculate the Riemannian gradient according to \eqref{obmgrad}
\REPEAT
\STATE Choose the step size $\delta_j$ according to \cite [p. 62]{absil2009optimization}
\STATE Find $\mathbf{u}_{j+1}$ by retraction in \eqref{retraction}

\STATE Update Riemannian gradient $\mathrm{grad}_{\mathbf{u}_{j+1}}f$ by using \eqref{obmgrad}
\STATE Calculate the vector transport $\mathcal{T}_{\mathbf{u}_j\rightarrow \mathbf{u}_{j+1}}(\bm{\mu}_{j})$ by using \eqref{transport}
\STATE Choose Polak-Ribi\`ere parameter $\alpha_j$ according to \cite [Eq. 8.24]{absil2009optimization}
\STATE Calculate conjugate search direction $\bm{\mu}_{j+1}$ by using \eqref{obmupdate}
\STATE Set $j=j+1$
\UNTIL $\left \| \mathrm{grad}_{\mathbf{u}_{j}}f \right \|\leq \varepsilon$
\STATE Set $\mathbf{\Phi}=\mathrm{diag}\big((\mathbf{u}^T_{j+1})^H\big)$
\end{algorithmic}
\end{algorithm}
\begin{algorithm}[t]
\caption{Alternating Optimization Algorithm}
\begin{algorithmic}[1]
\small
\STATE Set iteration index $t=1$, the initial point $\mathbf{u}^{(1)}$, convergence tolerance $\epsilon$, maximum iteration number $T_{\mathrm{max}}$.
\REPEAT
\STATE Solve \eqref{prob3} via \textbf{Algorithm 1} for given $\mathbf{u}^{(t)}$ and store the optimal solution $\mathbf{W}^{(t)}$ and $\mathbf{Z}^{(t)}$
\STATE Solve \eqref{prob5} via \textbf{Algorithm 2} for given $\mathbf{W}^{(t)}$ and $\mathbf{Z}^{(t)}$ and store the solution $\mathbf{u}^{(t+1)}$
\STATE Set $t=t+1$
\UNTIL $\left |f^{(t+1)}-f^{(t)}\right |\leq \epsilon$
\STATE Obtain the solution by $\mathbf{W}^*=\mathbf{W}^{(t)}$, $\mathbf{Z}^*=\mathbf{Z}^{(t)}$, and $\mathbf{u}^*=\mathbf{u}^{(t)}$
\end{algorithmic}
\end{algorithm}
\par
The proposed alternating optimization algorithm is summarized in \textbf{Algorithm 3}. Recall that the objective function is monotonically non-increasing after each iteration of both \textbf{Algorithm 1} and \textbf{Algorithm 2}. Therefore, the proposed alternating optimization algorithm is guaranteed to converge to a suboptimal solution of \eqref{prob2}.
\vspace*{2mm}
\section{Simulation Results}
\begin{table}[t]\caption{System Parameters}\vspace*{0mm}\label{tab:para} \centering
\begin{tabular}{|l|c|}\hline
\hspace*{-1mm}System bandwidth and carrier center frequency  & $200$ kHz and $2.4$ GHz \\
\hline
\hspace*{-1mm}Noise powers, $\sigma^2_{n_k}$ and $\sigma^2_{n_e}$&   \mbox{$-110$ dBm} \\
\hline
\hspace*{-1mm}BS maximum transmit power, $P_{\mathrm{max}}$ &  \mbox{$40$ dBm} \\
\hline
\hspace*{-1mm}Convergence tolerances, $\epsilon$ and $\varepsilon$ & \mbox{$10^{-3}$}\\
\hline
\end{tabular}\vspace*{-3mm}
\end{table}
\par
\begin{figure}[t]\vspace*{0mm}
\centering
\includegraphics[width=3.5 in]{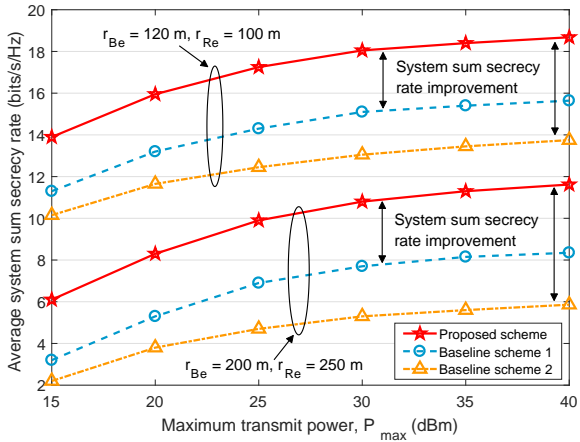} 
\vspace*{-1mm}
\caption{Average system sum secrecy rate (bits/s/Hz) versus maximum transmit power (dBm) with $K=3$, $N_{\mathrm{T}}=6$, and $M=6$.}\vspace*{-3mm} \label{fig:power_vs_SSR}
\end{figure}
We investigate the system performance of the proposed resource allocation scheme via simulations. Table \ref{tab:para} summarizes the parameters used in our simulation. In particular, the BS is at the center of a single cell with radius $500$ meters. One sector of the cell happens to be blocked by buildings and there are $K$ users randomly and uniformly distributed within this sector. An IRS is deployed to provide communication service for the users in this sector. We focus on the resource allocation design to achieve secure communication in this sector. Moreover, we also adopt two baseline schemes for comparison. For baseline scheme 1, we adopt an IRS with random phase $\phi_m$, $\forall m\in\mathcal{M}$ \cite{di2019smart}, and jointly optimize $\mathbf{w}_k$ and $\mathbf{Z}$. For baseline scheme 2, the BS does not generate AN (as in \cite{yu2019enabling} and \cite{chen2019intelligent}) and an IRS is employed for security provisioning. In this case, we jointly optimize only $\mathbf{w}_k$ and $\mathbf{\Phi}$ to achieve secure communication. $r_{\mathrm{Be}}$ and $r_{\mathrm{Re}}$ denote the distance from the BS to the eavesdropper and the distance from the IRS to the eavesdropper, respectively.
\par
In Figure \ref{fig:power_vs_SSR}, we study the average system sum secrecy rate versus the maximum transmit power. As expected, the system sum secrecy rates for the proposed scheme and the two baseline schemes increase monotonically with increasing $P_{\mathrm{max}}$. Moreover, we can see that the proposed scheme outperforms the baseline schemes. In fact, by jointly optimizing $\mathbf{\Phi}$, $\mathbf{w}_k$, and $\mathbf{Z}$, the proposed scheme can simultaneously facilitates a more favourable radio propagation environment for the users and impair the eavesdropper. In contrast, the two baseline schemes achieve significantly lower system sum secrecy rates, due to the random phase of the IRS for baseline scheme 1 and the lack of AN for baseline scheme 2. Besides, we can observe from Figure \ref{fig:power_vs_SSR} that the geometry of the network (i.e., the values of $r_{\mathrm{Be}}$ and $r_{\mathrm{Re}}$) has a significant impact on the system sum secrecy rate. This indicates that the location of the IRS needs to be chosen carefully for achieving the best possible system performance.
\par
\begin{figure}[t]\vspace*{0mm}
 \centering
\includegraphics[width=3.5 in]{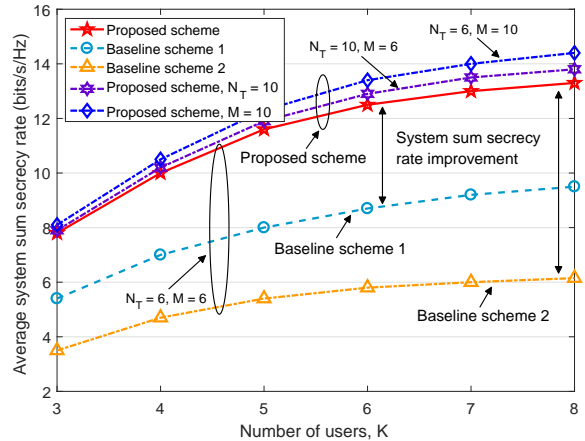}
\vspace*{-1mm}
\caption{Average system sum secrecy secrecy rate (bits/s/Hz) versus number of users with $P_{\mathrm{max}}=20$ dBm, $r_{\mathrm{Be}}=200$ m, and $r_{\mathrm{Re}}=250$ m.}\vspace*{-3mm}\label{fig:userk_vs_SSR}
\end{figure}

Figure \ref{fig:userk_vs_SSR} shows the average system sum secrecy rate versus the number of legitimate users with $P_{\mathrm{max}}=20$ dBm, $N_{\mathrm{T}}=6$, and $M=6$. We observe that the system sum secrecy rates achieved by the proposed scheme and the two baseline schemes monotonically increase with $K$. This is due to the fact that both the proposed scheme and the two baseline schemes are able to exploit multiuser diversity. To investigate the performance gain attained by deploying IRSs, we show the system sum secrecy rate of the proposed scheme for two additional cases: Case 1 with $N_{\mathrm{T}}=10$ and $M=6$ and Case 2 with $N_{\mathrm{T}}=6$ and $M=10$. We observe that Case 2 results in a larger performance gain over the system with the default parameters ($N_{\mathrm{T}}=6$ and $M=6$) compared to Case 1. The reasons behind this are two-fold. On the one hand, the extra phase shifters can reflect more power of the signal received from the BS which leads to a power gain. On the other hand, they also provide higher flexibility in resource allocation which improves the beamforming gain for the IRS-user links.   
\vspace*{-2mm}
\section{Conclusion}
\vspace*{-2mm}
In this paper, we proposed an efficient resource allocation algorithm to achieve secure communication in IRS-assisted multiuser MISO systems. AN is injected by the BS to enhance physical layer security. Due to the non-convexity of the formulated optimization problem, we developed an alternating optimization algorithm with guaranteed convergence. Our simulation results reveal that the proposed scheme can significantly enhance the security of IRS-assisted wireless communication systems compared to the two baseline schemes, which respectively do not optimize the IRS phase shift matrix or do not exploit AN.
\section*{Appendix- Proof of Theorem 1}
\vspace*{-2mm}
We note that if $R_k-C_k^{\mathrm{E}}\leq0$, the proposed algorithm would stop transmitting information to user $k$ and allocate the corresponding power to other users. In this case, the optimal beamforming vector for user $k$ is $\mathbf{w}^*_k=\mathbf{0}$ which implies $\mathrm{Rank}(\mathbf{W}^*_k)=0$. Next, for the case where $P_{\mathrm{max}}>0$ and $R_k-C_k^{\mathrm{E}}>0$, we show that the optimal beamforming matrix $\mathbf{W}_k^*$ is indeed a rank-one matrix. To start with, we rewrite \eqref{prob4} in the following equivalent form:
\vspace*{-2mm}
\begin{eqnarray}
\label{prob6}
&&\hspace*{0mm}\underset{\mathbf{W}_k,\mathbf{Z}\in\mathbb{H}^{\mathit{N}_{\mathrm{T}}},\eta,\tau_k,\iota}{\mino} \,\, \,\,\eta \\
\mbox{s.t.}\hspace*{0mm}
&&\hspace*{-4mm}\mbox{C1},\mbox{C3},\mbox{C4},~\mbox{C6: }\overline{F_1}+\overline{F_2}-\widetilde{G_1}-\widetilde{G_2}\leq\eta,\notag
\\
&&\hspace*{-4mm}\mbox{C7: }\tau_k\hspace*{-0.5mm}\leq\hspace*{-0.5mm} \underset{r\in\mathcal{K}}{\sum }\mathrm{Tr}(\mathbf{W}_r\mathbf{G}_k^H\mathbf{u}\mathbf{u}^H\mathbf{G}_k)\hspace*{-0.5mm}+\hspace*{-0.5mm}\mathrm{Tr}(\mathbf{Z}\mathbf{G}_k^H\mathbf{u}\mathbf{u}^H\mathbf{G}_k),\notag\\
&&\hspace*{-4mm}\mbox{C8: }\iota\leq\mathrm{Tr}(\mathbf{Z}\mathbf{L}^H\mathbf{u}\mathbf{u}^H\mathbf{L})\notag,\\[-6mm]\notag
\end{eqnarray}
where $\overline{F_1}\hspace*{-0.5mm}=\hspace*{-0.5mm}-\underset{k\in\mathcal{K}}{\sum }\mathrm{log}_2(\tau_k+\sigma^2_{n_k})$ and $\overline{F_2}\hspace*{-0.5mm}=\hspace*{-0.5mm}-\underset{k\in\mathcal{K}}{\sum }\mathrm{log}_2(\iota+\sigma^2_{n_e})$, and $\tau_k$ and $\iota$ are auxiliary optimization variables.
\par
Problem \eqref{prob6} is jointly convex with respect to all optimization variables. Moreover, it can be verified that Slater's
condition holds \cite{boyd2004convex}. Therefore, strong duality holds, i.e., we can obtain the optimal solution of \eqref{prob6} by solving the dual problem \cite{boyd2004convex}. The Lagrangian function of \eqref{prob6} in terms of beamforming matrix $\mathbf{W}_\mathit{k}$ is given by
\vspace*{-1mm}
\begin{eqnarray}
\label{Lagrangian}
\mathcal{L}\hspace*{-1mm}&&\hspace*{-6mm}=\xi\underset{k\in\mathcal{K}}{\sum }\mathrm{Tr}(\mathbf{W}_k)-\underset{ k\in\mathcal{K}}{\sum}\mathrm{Tr}(\mathbf{W}_\mathit{k}\mathbf{Y}_\mathit{k})\notag\\
&&\hspace*{-6mm}-\kappa\mathrm{Tr}\Big(\hspace*{-0.5mm}\big[\nabla_{\mathbf{W}}G_1(\mathbf{W}^i,\mathbf{Z}^i)+\nabla_{\mathbf{W}}G_2(\mathbf{W}^i,\mathbf{Z}^i)\big]^H(\mathbf{W}\hspace*{-0.5mm}-\hspace*{-0.5mm}\mathbf{W}^i)\Big)\notag\\
&&\hspace*{-6mm}-\underset{k\in\mathcal{K}}{\sum }\lambda_k\underset{r\in\mathcal{K}}{\sum }\mathrm{Tr}(\mathbf{W}_r\mathbf{G}_k^H\mathbf{u}\mathbf{u}^H\mathbf{G}_k)+\Upsilon,\\[-4mm]\notag
\end{eqnarray}
where $\Upsilon$ denotes the collection of the optimization variables of the primal and dual problems and constant terms that are not relevant to the proof. $\xi$, $\kappa$, and $\lambda_k$ denote the scalar Lagrange multipliers associated with constraints $\mbox{C1}$, $\mbox{C6}$, and $\mbox{C7}$. $ \mathbf{Y}_k\in \mathbb{C}^{N_\mathrm{T}\times N_\mathrm{T}}$ is the Lagrange multiplier matrix associated with constraint $\mbox{C4}$. The dual problem of \eqref{prob4} is given by
\vspace*{-1mm}
\begin{equation}
\underset{\substack{\mathbf{Y}_k\succeq \mathbf{0},\\\xi,\kappa,\lambda_k\geq0~}}{\mathrm{maximize}~~} \underset{\substack{\mathbf{W}_k,\mathbf{Z}\in\mathbb{H}^{\mathit{N}_{\mathrm{T}}},\\\eta,\tau_k,\iota}}{\mathrm{minimize}}~~ \mathcal{L}(\mathbf{W}_k,\mathbf{Z},\eta,\mathbf{Y}_k,\xi,\kappa,\lambda_k).\label{DP}\\[-1mm]
\end{equation}
Then, we investigate the structure of the optimal $\mathbf{W}^*_k$ of dual problem \eqref{prob4} by applying the Karush-Kuhn-Tucker (KKT) conditions. In particular, the KKT conditions associated with $\mathbf{W}_k^*$ are as follows
\vspace*{-2mm}
\begin{equation}
\mathrm{K1\hspace*{-1mm}:}\xi^*,\kappa^*,\lambda_k^*\geq 0, \mathbf{Y}_k^*\succeq \mathbf{0},~
\mathrm{K2\hspace*{-1mm}:}\mathbf{Y}_k^*\mathbf{W}_k^*=\mathbf{0},~
\mathrm{K3\hspace*{-1mm}:}\triangledown_{\mathbf{W}_k^*}\mathcal{L}=\mathbf{0}, \\[-1mm]
\end{equation}
where $\xi^*$, $\kappa^*$, $\lambda_k^*$, and $\mathbf{Y}_k^*$ denote the optimal Lagrange multipliers for dual problem ($\ref{DP}$), and $\triangledown_{\mathbf{W}_k^*}\mathcal{L}$ represents the gradient vector of \eqref{Lagrangian} with respect to $\mathbf{W}_k^*$. To facilitate the proof, we rewrite $\mathrm{K3}$ explicitly as follows
\vspace*{-2mm}
\begin{equation}
    \mathbf{Y}_k^*=\xi^*\mathbf{I}_{N_{\mathrm{T}}}-\mathbf{\Delta},\label{alK3}\\[-2mm]
\end{equation} where $\mathbf{\Delta}$ is given by
\vspace*{-1mm}
\begin{eqnarray}
\mathbf{\Delta}&&\hspace*{-5mm}=\kappa^*\Big(\hspace*{-0.5mm}\nabla_{\mathbf{W}}G_1(\mathbf{W}^i,\mathbf{Z}^i)+\nabla_{\mathbf{W}}G_2(\mathbf{W}^i,\mathbf{Z}^i)\Big)\notag\\&&\hspace*{-5mm}+\underset{r\in\mathcal{K}}{\sum }\lambda^*\mathbf{G}_r^H\mathbf{u}\mathbf{u}^H\mathbf{G}_r.\\[-5mm]\notag
\end{eqnarray}
\par
Next, by revealing the structure of matrix $\mathbf{Y}_k^*$, we prove that the optimal beamforming matrix $\mathbf{W}^*$ is indeed a rank-one matrix. To start with, we first denote the maximum eigenvalue of matrix $\mathbf{\Delta}$ as $\nu ^{\mathrm{max}}_{\mathbf{\Delta}}\in\mathbb{R}$. We note
that the cases where multiple eigenvalues have the same value
$\nu ^{\mathrm{max}}_{\mathbf{\Delta}}$ and where $\mathbf{\Delta}$ is negative-semidefinite matrix occur with probability zero, due to the randomness of
the channels. Reviewing \eqref{alK3}, if $\nu ^{\mathrm{max}}_{\mathbf{\Delta}}>\xi^*$, then $\mathbf{Y}_k^*$ cannot be a positive semidefinite matrix which contradicts K1. On the other hand, if $\nu ^{\mathrm{max}}_{\mathbf{\Delta}}<\xi^*$, then $\mathbf{Y}_k^*$ must be a positive definite matrix with full rank. In this case, considering K2, $\mathbf{W}_k^*$ is forced to be $\mathbf{0}$ which is obviously not the optimal solution for $P_{\mathrm{max}}>0$ and $R_k-C_k^{\mathrm{E}}>0$.  In addition, we note that there exists at least one optimal solution with $\xi^*>0$ such that constraint C1 is met with equality. Therefore, for the optimal solution, the equality $\nu ^{\mathrm{max}}_{\mathbf{\Delta}}=\xi^*$ must hold 
which results in $\mathrm{Rank}(\mathbf{Y}_k^*)=N_\mathrm{T}-1$. Next, we construct a bounded optimal solution based on the above discussion. In particular, we construct a unit-norm vector $\mathbf{e}_{\mathbf{\Delta}}^{\mathrm{max}}\in \mathbb{C}^{N_{\mathrm{T}}}$ which lies in the null space of $\mathbf{Y}_k^*$, i.e., $\mathbf{Y}_k^*\mathbf{e}_{\mathbf{\Delta}}^{\mathrm{max}}=\mathbf{0}$. We note that $\mathbf{e}_{\mathbf{\Delta}}^{\mathrm{max}}$ denotes the eigenvector of matrix $\mathbf{\Delta}$ corresponding to the maximum eigenvalue $\nu ^{\mathrm{max}}_{\mathbf{\Delta}}$ with unit norm. Therefore, for $P_{\mathrm{max}}>0$ and $R_k-C_k^{\mathrm{E}}>0$, the optimal beamforming matrix $\mathbf{W}_k^*$ is indeed a rank-one matrix which can be expressed as $\mathbf{W}_k^*= \zeta \mathbf{e}_{\mathbf{\Delta}}^{\mathrm{max}}(\mathbf{e}_{\mathbf{\Delta}}^{\mathrm{max}})^H$, where $\zeta$ is a parameter to adjust $\mathbf{W}_k^*$ such that constraint $\mbox{C1}$ is satisfied with equality. \qed

\vspace*{-1mm}
\bibliographystyle{IEEEtran}
\bibliography{Reference_list}
\end{document}